\documentclass[useAMS,usenatbib]{mn2e}
\usepackage{graphicx,bm}

\title[Very Large and Very Small]{Connecting the Very Large and the Very
Small: \\ Effective Particle Mass in Curved-space \\ Cosmological Models}

\author[Peter R. Phillips]{Peter R. Phillips \\
Department of Physics, Washington University, St. Louis, MO 63130 }

\begin{document}

\maketitle

\begin{abstract}
We investigate the propagation of electromagnetic fields and
potentials in the plasma of the early Universe, assuming a
Friedmann-Robertson-Walker background with negative curvature. Taking over
results from classical plasma physics, we show that charged particles will
acquire an effective mass that has not only the expected thermal component
but also a non-thermal component due to the influence of distant matter.
Although this is a direct effect of the vector potential, we show the
theory is nevertheless gauge invariant. This phenomenon is therefore in
the same category as the Aharonov-Bohm effect. The non-thermal component
becomes increasingly important with time, and in some cosmological models
can prove to be of decisive importance in bringing about the phase
transition that generates normal masses.
\end{abstract}




\section{INTRODUCTION}
\label{sec:intro}

In this paper we study the effective mass of charged particles in a plasma,
with particular emphasis on the early Universe, before ordinary masses are
generated. In Minkowski space, particles of mass $m$ acquire an
additional effective thermal mass, $m_{\mathrm{th}}$, of general order of
the temperature of the plasma. This is normally derived using
path-integral methods \citep{nar2}. We will estimate the
effective mass from a different point of view, which will allow a
straightforward extension to an expanding Friedmann-Robertson-Walker
(FRW) space.

Our main conclusion is that if the FRW space has negative curvature, the
effective mass acquires a new non-thermal component, $m_{\mathrm{nt}}$,
in addition to the thermal one. This new component is due to the cumulative
effect of distant matter; it is not constant, but increases with time
relative to the thermal component. In today's standard cosmology, where
inflation leads to a flat space, $m_{\mathrm{nt}}$ will not appear. But
there are alternative cosmological models in which it could be of
decisive importance.

$m_{\mathrm{nt}}$ appears because in curved space the potentials (but not
the fields) have a tail \citep{dew,nar1,hag1}, as shown in figures
\ref{fig:HPEPS} and \ref{fig:ATEPS}. In the tail, we have potentials but
no fields, so $m_{\mathrm{nt}}$ is a direct effect of the potentials. This
is a quantum mechanical result, though we have derived it in the context
of classical gravity. As with the Aharonov-Bohm effect \citep{ahar,tono},
the appearance of $m_{\mathrm{nt}}$ at first seems to violate gauge
invariance, and we have to show why this is actually not so.

\section{NOTATION}
\label{sec:notation}

We consider a model consisting of a single massless, charged scalar field,
$\Phi$, interacting with the usual electromagnetic fields and potentials.
Units are chosen so that $c = \hbar = 1$. Our sign conventions are
those of \citet{wein2}, so the metric signature in Minkowski
space is $( - + + + )$. When gravitational effects are introduced,
$g = -{\mathrm{Det}}\,(g_{\mu\nu})$.

For the early Universe we assume a FRW metric corresponding to
negative curvature. (Positive curvature would give similar effects, but
involves an awkward discrete distribution of wave numbers.) We use
coordinates $t$, $\chi$, $\theta$, $\phi$, labeled 0, 1, 2, 3,
so that
\begin{eqnarray}
{\mathrm d} s^2  & = & -{\mathrm d} t^2 +
R^2 (t) \left( {\mathrm d} \chi^2 + \sinh^2 \chi \, {\mathrm d} \theta^2 \right.
\nonumber \\
  & & \left. {} + \sinh^2 \chi \sin^2 \theta \, {\mathrm d} \phi^2 \right) ,
\label{eq:ds1} \\
g_{\mu\nu}  & = & {\mathrm{diag}} \left( -1, R^2 (t),
R^2 (t) \sinh^2 \chi, \right. \nonumber \\
  & & \left.  R^2 (t) \sinh^2 \chi \sin^2 \theta \right) ,
\label{eq:gmetric1} \\
\sqrt{g}  & = & R^3 (t) \sinh^2 \chi \sin \theta  . \label{eq:gdet1} 
\end{eqnarray}
The expansion parameter, $R(t)$, has the dimension of length, and can be
thought of as the radius of the Universe at time $t$.

The conformal time, $\eta$, is related to the ordinary time $t$ by
${\mathrm d} \eta = {\mathrm d} t/R(t)$. In terms of $\eta$, $\chi$, $\theta$,
$\phi$ (all dimensionless) we have:
\begin{eqnarray}
{\mathrm d} s^2  & = & R^2 (\eta)\left(-{\mathrm d}  \eta^2 +
{\mathrm d} \chi^2 + \sinh^2 \chi \, {\mathrm d} \theta^2 \right. \nonumber \\
  & & \left. {} + \sinh^2 \chi \sin^2 \theta \, {\mathrm d} \phi^2 \right) ,
\label{eq:ds2} \\
g_{\mu\nu}  & = & R^2 (\eta){\mathrm{diag}} \left( -1, 1, \sinh^2 \chi,
\sinh^2 \chi \sin^2 \theta \right) , \label{eq:gmetric2} \\
\sqrt{g}  & = & R^4 (\eta) \sinh^2 \chi \sin \theta . \label{eq:gdet2} 
\end{eqnarray}

Maxwell's equations in free space are:
\begin{eqnarray}
\left( \sqrt{g} F^{\alpha \beta} \right) _{,\beta}  & = & 0 ,
\label{eq:max1} \\
F_{\alpha \beta , \gamma} + F_{\beta \gamma , \alpha} 
+ F_{\gamma \alpha , \beta}  & = & 0 . \label{eq:max2}
\end{eqnarray}

Potentials and a gauge condition will be introduced later.

\section{EFFECTIVE THERMAL MASS IN A \\
PLASMA IN MINKOWSKI SPACE}
\label{sec:Mink}

We start from the simple, and at first sight pointless,
observation that the kinetic-energy term of the Klein-Gordon equation for
a charged scalar field in Minkowski space,
$\Phi^{*} ({\mathbf p} - q{\mathbf A})^2 \Phi$, when expanded, gives the
term $\Phi^{*} q^2 {\mathbf A}^2 \Phi$. This has the same sign as the mass
term $\Phi^{*} m^2 \Phi$, suggesting that in some circumstances, such as
in a plasma, $\langle q^2 {\mathbf A}^2 \rangle$ might play the role of
the square of an effective thermal mass. $\langle {\mathbf A}^2 \rangle$ is
the sum of contributions from particles in the immediate neighborhood,
roughly, those within the Debye sphere \citep{stur}, and
should therefore be written $\langle (\sum_i {\mathbf A_i})^2 \rangle$.
When we sum over all contributions, the ${\mathbf A_i}$ potentials will be
randomly oriented, and will add according to the theory of random flights
\citep{hugh1}. $\langle \sum_i {\mathbf A_i} \rangle$ will then be zero,
but $\langle (\sum_i {\mathbf A_i})^2 \rangle$ will not.

In a proper quantum mechanical treatment we have to represent the source
particles by means of a density matrix. This is not important for a plasma
in Minkowski space, but will be essential when we come to curved space, so
we introduce the idea here. The density matrix provides a way of including
all possible configurations of source particles, which are not, of course,
individually observed. Instead of ${\mathbf A_i}$ we should write
${\mathbf A_{\mu i}}$, where $\mu$ specifies a particular configuration.
$\langle {\mathbf A}^2 \rangle$ is then a double sum, over the particles
within a configuration, and over the various possible configurations; up
to a normalization factor,
$\langle \sum_{\mu} ( \sum_i {\mathbf A_{\mu i}})^2 \rangle$.
Dimensional considerations indicate that in the conditions of the early
Universe this should have a value of order $T^2$, where $T$ is the
temperature of the plasma.

We are free to make an overall gauge transformation with an arbitrary
function $\psi (x^{\mu})$. The various ${\mathbf A_i}$ become 
${\mathbf A_i} + {\mathbf \nabla} \psi$, and the source particles acquire
corresponding phases. But we should note that this freedom is not the same
as freedom to choose any gauge we wish to describe the propagation of the
potentials. For example, we might decide to switch from Lorenz gauge
\citep{jack2} to Coulomb gauge. This would introduce a separate gauge
function for each ${\mathbf A_{\mu i}}$, and there would be no single
gauge function that could be used to generate a compensating phase for any
chosen interacting particle. If we are concerned with the propagation of
potentials we have to choose a gauge from the beginning that is suitable
for that purpose, as discussed in section \ref{sec:dipot}. This choice is
not altered by an overall gauge transformation.

We can argue, of course, that in Minkowski space we are not really dealing
with a direct effect of the vector potential, because electric and magnetic
fields are also present. and they are the actual physical agents. This
is true, but, as we will see in section \ref{sec:FRW}, it is no longer true
in a curved space, where we have potentials without fields.

We will need to know the potential due to a single particle in the plasma at
distances larger than the Debye length. If the particle is stationary, the
potential declines exponentially. Surprisingly, however,
starting in the 1950's several authors \citep{neuf,tapp1,mont1} discovered
that a test charge \textit{moving uniformly} is not
exponentially screened, but generates the field of a quadrupole in a
collisionless plasma. The restriction to a collisionless plasma was lifted
in later papers \citep{yu,schr}, where it was shown that the far
field of a moving test charge is that of a dipole with strength of order
$q\tau_t V$, where $q$ is the charge, $\tau_t$ the collision time as
measured by $t$, and $V$ the velocity. This dipole form ensures that in
Minkowski space the contribution of particles to the effective mass falls
off rapidly with distance beyond the Debye radius. We shall see, however,
that this is no longer the case in a curved space.
 
\section{SOURCE PARTICLES IN THE PLASMA}
\label{sec:source}

To study the propagation of fields and potentials in a curved FRW space,
we first need a model of the motion of the source particles that produce
them. We are concerned with the early Universe, before
any symmetry-breaking transitions that may generate particle masses as we
know them. The properties of the plasma are therefore determined by a single
number, the temperature. The photons will have the familiar Planck
distribution. For simplicity, we assume the charged particles are associated
with a single massless scalar field. This field will have an effective
thermal mass, $m_{\mathrm{th}}$, due to its interaction with the photons,
so that $m_{\mathrm{th}} \approx T$. The charged particles will have a
thermal distribution appropriate to this mass and temperature. The plasma
will therefore be weakly relativistic; for simplicity, however, we will
carry over some results from non-relativistic plasmas. Other plasma
parameters are determined by $T$; for example, the plasma frequency,
$\omega_p$, will be of order $T$.

A charge that is part of the plasma, rather than being a test charge
constrained to move in a certain way, will itself experience collisions. We
can picture such a particle as describing Brownian motion, with its
screening cloud sometimes closer, sometimes farther, but never fully
established. We will model the most important aspect of a single step of
this process by setting up a local set of Cartesian coordinates and placing
a stationary screening charge $-q$ at the origin. A charge $q$ is initially
at rest on the $z$ axis at $z = -V \tau_t$, then begins to move with uniform
velocity $V$ along the $z$ axis, starting at $t = -\tau_t$ and ending at
$t = \tau_t$, when it again comes to rest. Note that this choice of origin
for $t$ differs from that of the cosmic time used in section
\ref{sec:notation}; $\eta$, defined below, will differ in a similar way. We
reconcile these differences later, in section \ref{sec:mass_1}.

We are now going to take the result for a uniformly moving test charge and
use it to calculate the field around our model charge that is subject
to Brownian motion. We justify this step as follows. Imagine that the test
charge, instead of moving uniformly from an indefinite time in the past,
is actually stationary until time $t=0$, and then starts moving uniformly.
There will be transient fields, but after a few collision times the final
field of a moving dipole will be established. During the transient period,
the test charge will be moving away from its screening charge, which only
gradually picks up speed and trails along behind. This is similar to the
motion of the charge in our model. Since the transient fields form a
bridge between the initial and final fields they must have a similar form
to the field of the test charge, and extend comparably far.

During the period that the charge is moving (the \textit{pulse}), the dipole
moment will be $\overline{D} (t) = qVt$. Before and after the pulse the
dipole moment remains at a constant value, but this is of little interest
because a constant dipole moment generates no magnetic field or vector
potential. It is well known that Maxwell's equations separate in conformal
time, so we will write the dipole moment in terms of $\eta$. Define
$\tau = \tau_t/R_{\mathrm{s}}(t)$ (dimensionless), so that the pulse extends
from $\eta = -\tau$ to $\eta = \tau$. The subscript `s' on $R_{\mathrm{s}}$
indicates the source time. $R_{\mathrm{s}}(t)$ can be treated as constant
during the pulse, and the dipole moment can be expressed as a Fourier
integral:
\begin{eqnarray}
\overline{D} (\eta)  & = & \frac{1}{2\pi} \int D(n) \exp (-{\mathrm i} n\eta)
\, {\mathrm d} n , \label{eq:dipfour} \\ 
D(n)  & = & \frac{2{\mathrm i} qVR_{\mathrm{s}} \sin (n\tau)}{n^2} . 
\label{eq:dipstrength}
\end{eqnarray}

We estimate the value of $\tau_t$ in section \ref{sec:mass_1}. The $z$ axis
of the local system used in this section will be parallel to the polar axis
of the polar coordinates used in the bulk of the paper.

\section{DIPOLE FIELDS}
\label{sec:dipfield}

We will need only the three fields $F_{10}$, $F_{20}$ and $F_{12}$.
Since Maxwell's equations separate in conformal time the elementary
solutions can be written
\begin{eqnarray}
F_{10}  & = & f_{10} (\chi, \theta, \phi) \exp(-{\mathrm i} n\eta) ,
\label{eq:F10def} \\
F_{20}  & = & f_{20} (\chi, \theta, \phi) \exp(-{\mathrm i} n\eta) ,
\label{eq:F20def} \\
F_{12}  & = & f_{12} (\chi, \theta, \phi) \exp(-{\mathrm i} n\eta) .
\label{eq:F12def}
\end{eqnarray}

For dipole fields we try the following forms:
\begin{eqnarray}
f_{10}  & = & f_1 (\chi) P_1 (\theta) , \label{eq:f10} \\
f_{20}  & = & f_2 (\chi) N_1 (\theta) , \label{eq:f20} \\
f_{12}  & = & f_3 (\chi) N_1 (\theta) , \label{eq:f12}
\end{eqnarray}
with angular functions $P_1 (\theta) = \cos \theta$ and 
$N_1 (\theta) \equiv {\mathrm d}  P_1 (\theta)/{\mathrm d}  \theta
= -\sin \theta$.

Maxwell's equations now give (with a prime meaning
${\mathrm d} /{\mathrm d} \chi$):
\begin{eqnarray}
{\mathrm i} n \sinh^2 \chi f_1 - 2 f_3  & = & 0 , \label{eq:dip1} \\
{\mathrm i} n f_2 - f_3^{\,\prime}  & = & 0 , \label{eq:dip2} \\
f_1 - f_2^{\,\prime} + {\mathrm i} n f_3  & = & 0 . \label{eq:dip3} 
\end{eqnarray}

\subsection{\label{subsec:magnetic}Dipole Magnetic Field}

From (\ref{eq:dip1}), (\ref{eq:dip2}) and (\ref{eq:dip3}) we obtain the
equation for $f_3$ alone
\begin{equation}
\frac{{\mathrm d} ^2 f_3}{{\mathrm d}  \chi^2} +
\left(n^2 - \frac{2}{\sinh^2 \chi} \right) f_3 = 0 . \label{eq:f3}
\end{equation}

This is the analog of equation 16 of \citet{mash1}, which he
derived for a space of positive curvature. Define
$f_4 (\chi) = f_3 (\chi) / \sinh \chi$. Then $f_4$ satisfies
\begin{equation}
\frac{{\mathrm d} ^2 f_4}{{\mathrm d}  \chi^2} +
2 \coth \chi \frac{{\mathrm d}  f_4}{{\mathrm d}  \chi} +
\left(n^2 + 1 - \frac{2}{\sinh^2 \chi} \right) f_4 = 0 . \label{eq:f4}
\end{equation}
This is a special case, for $l=1$ or $l=-2$, of the equation for
hyperbolic spherical functions, with solution \citep{buch1,band1}:
\begin{equation}
f_4 (l, \chi) = N_l (-1)^{l+1} \sinh^{l} \chi
\frac{{\mathrm d} ^{l+1}}{{\mathrm d}  (\cosh \chi)^{l+1}} \cos (n\chi)
\label{eq:rod}
\end{equation}
where $N_l$ is a normalization factor that is irrelevant here. Choosing
$l=1\;(-2)$ we get the solutions of (\ref{eq:f4}) that are regular
(singular) at $\chi = 0$. With $l+1 < 0$ interpreted as integration:
\begin{eqnarray}
f_{4,{\mathrm{reg}}}  \equiv f_4 (1, \chi)
& = & \frac{1}{\sinh^2 \chi}
[n \sinh \chi \cos(n \chi) \nonumber \\
  & & {} - \cosh \chi \sin (n \chi)] \label{eq:f4reg} \\
f_{4,{\mathrm{sing}}}  \equiv f_4 (-2, \chi)
& = & \frac{1}{\sinh^2 \chi}
[n \sinh \chi \sin (n \chi) \nonumber \\
  & & {} + \cosh \chi \cos (n \chi )] \label{eq:f4sing}
\end{eqnarray}

$f_3 $ is constructed from that combination of $f_{4,{\mathrm{reg}}}$ and
$f_{4,{\mathrm{sing}}}$ that is proportional to $\exp (i n \chi)$,
because when combined with $\exp (-i n \eta)$ this represents
outgoing waves:
\begin{equation}
f_3 =  \frac{C_1 (n) \exp ({\mathrm i} n\chi)}{\sinh \chi}
\left( \cosh \chi - {\mathrm i} n \sinh \chi \right) , \label{eq:f3out}
\end{equation}
where $C_1 (n)$ is a normalizing factor to be determined.

It is convenient at this point to express $f_3$ in terms of
$u=\tanh (\chi/2)$:
\begin{equation}
f_3 = \frac{C_1 (n) \exp ({\mathrm i} n\chi)}{2 u}
\left( 1 - 2{\mathrm i} nu + u^2 \right) . \label{eq:f3out2}
\end{equation}

\subsection{\label{subsec:electric}Dipole Electric Fields}

From (\ref{eq:dip1}) and (\ref{eq:f3out}) we derive the
equation for the radial electric dipole field:
\begin{equation}
f_1 = \frac{-{\mathrm i}  C_1 (n) \exp({\mathrm i} n \chi )
(1-u^2)^2}{4nu^3} \left( 1 - 2{\mathrm i} nu + u^2 \right) .
\label{eq:f1out}
\end{equation}

Similarly, from (\ref{eq:dip2}) and (\ref{eq:f3out2}) we derive the
equation for the transverse electric field:
\begin{eqnarray}
f_2 & = & \frac{{\mathrm i} C_1 (n) \exp({\mathrm i} n\chi)}{4nu^2} \times
\nonumber \\`
  & & \left[ (1-u^2)^2 - 2{\mathrm i} nu(1+u^2) - 4n^2u^2 \right] .
\label{eq:f2out}
\end{eqnarray}

\subsection{\label{subsec:normfield}Normalizing the Fields}

From (\ref{eq:dipfour}) and (\ref{eq:dipstrength}) we can get the
near-field expression for the radial component of $E$, and by comparison
with (\ref{eq:f1out}) arrive at the form of the normalizing factor
$C_1(n)$. On the polar axis, at small distances, the radial ${\mathbf E}$
field at frequency $n$ is
\begin{equation}
E_r = -\frac{\partial}{\partial r}
\left(\frac{D(n) \exp(-{\mathrm i} n\eta)}{r^2} \right)
= \frac{2D(n) \exp(-{\mathrm i} n\eta)}{r^3} , \label{eq:Er_near}
\end{equation}
where $r=R_{\mathrm{s}} \chi$.

In the same limit (small $\chi$), (\ref{eq:f1out}) gives
$f_1 (\chi) = -2{\mathrm i}  C_1 (n) /(n \chi^3)$, and so, on the axis,
\begin{equation}
F_{10} (\eta, \chi) = -2{\mathrm i}  C_1 (n) R_{\mathrm{s}}^3
\exp(-{\mathrm i}  n \eta)/(n r^3) .
\end{equation}

We now transform to a local Minkowski frame with coordinates $t$, $r$:
\begin{eqnarray}
\overline{F}_{10} (t, r) & = & \frac{\partial \eta}{\partial t} 
\frac{\partial \chi}{\partial r} F_{10} (\eta, \chi) \nonumber \\
 & = & -2{\mathrm i}  C_1 (n) R_{\mathrm{s}} \exp(-{\mathrm i}  n \eta)/(n r^3) .
\label{eq:F10bar}
\end{eqnarray}

$\overline{F}_{10}$ is just the conventional radial electric field as given
in (\ref{eq:Er_near}), so using (\ref{eq:dipstrength}):
\begin{equation}
C_1 (n) = \frac{-2qV \sin(n\tau)}{n} .
\label{eq:normC1}
\end{equation}

\section{PROPAGATION OF THE MAGNETIC \\
FIELD}
\label{sec:propmag}

(\ref{eq:f3out2}), (\ref{eq:normC1}), (\ref{eq:f12}) and (\ref{eq:F12def})
give the Fourier transform of the magnetic field:
\begin{eqnarray}
F_{12} & = & \frac{-qV \sin (n \tau) N_1 (\theta)
\exp [{\mathrm i} n(\chi - \eta)]}{nu} \times \nonumber \\
 & & \left( 1 - 2{\mathrm i} nu + u^2 \right) . \label{eq:F12out}
\end{eqnarray}

\begin{figure}
\centering
\includegraphics[scale=0.5]{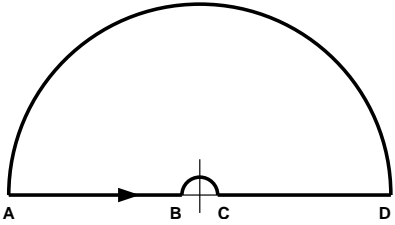}
\caption{\label{fig:contour1} Typical contour for Fourier synthesis of
magnetic and electric fields.  }
\end{figure}

We transform back into $\chi$, $\eta$ space by dividing by $2\pi$ and
integrating over $n$ along the real axis. For $\chi > \eta + \tau$ both
exponentials in $\sin (n \tau)$ allow us to close in the UHP, using the
contour of figure \ref{fig:contour1}. $\sin (n\tau)/n$ is regular at $n=0$,
so we get zero, as required by causality. Similarly, when
$\chi < \eta - \tau$ we can close in the LHP, using a contour that is the
inverse of figure \ref{fig:contour1}, and we again get zero.

For $\eta - \tau < \chi < \eta + \tau$ we must write $\sin (n \tau)
= (\exp({\mathrm i} n \tau) - \exp(-{\mathrm i} n \tau))/2{\mathrm i} $,
divide the integrand into two pieces, and close the first integral in the
UHP and the second in the LHP. Combining these two integrals we get
\begin{equation}
F_{12} = \frac{qV \sin \theta (1+u^2)}{2u} . 
\label{eq:F12out2}
\end{equation}

\begin{figure}
\centering
\includegraphics[scale=0.5]{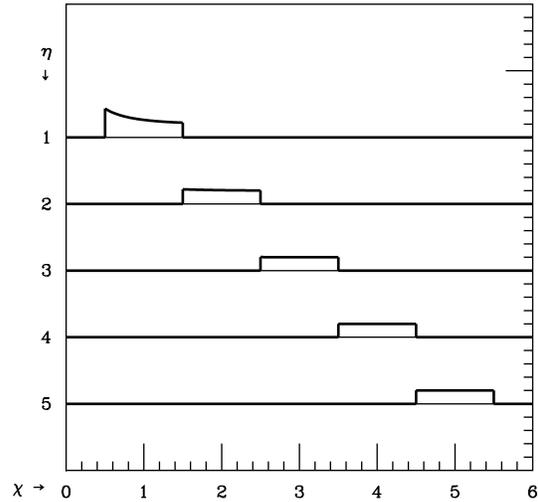}
\caption{\label{fig:HPEPS} Propagation of the magnetic field, $F_{12}$:
snapshots of the pulse for various values of conformal time, $\eta$. For
each $\eta$, the pulse is confined to the range
$\eta - \tau < \chi < \eta + \tau$; within that range we plot
$(1+u^2)/u$. To render the pulse visible on the graph, we have
arbitrarily set $\tau = 0.5$, far larger than it is in practice.  }
\end{figure}

The propagation of $F_{12}$ is shown in figure \ref{fig:HPEPS}. An
unexpected feature of the pulse is that the amplitude does not tend to zero
for large $\chi$, but to a constant asymptotic value. The
\textit{conventional} ${\mathbf H}$ field, does, of course, tend to zero,
in fact exponentially. We can see this at a particular instant by setting up
local Cartesian coordinates at a particular observation point, P, at rest in
the FRW space. Choose a point in the equatorial plane, $\theta = \pi/2$,
with azimuth $\phi=0$. Choose ${\mathrm d} x$ parallel to ${\mathrm d} \chi$,
${\mathrm d} y$ parallel to ${\mathrm d} \theta$, and ${\mathrm d} z$ parallel
to ${\mathrm d} \phi$. Then ${\mathrm d} x = R_{\mathrm{p}} {\mathrm d} \chi$,
and ${\mathrm d} y = R_{\mathrm{p}} \sinh \chi {\mathrm d}  \theta$, where the
subscript 'p' refers to the particular instant chosen. The conventional
$H_z$ field is $\overline{F_{12}}$, given by
\begin{eqnarray}
\overline{F_{12}} & = & \frac{\partial \chi}{\partial x}
\frac{\partial \theta}{\partial y} F_{12} \nonumber \\
  & = & F_{12}/(R_{\mathrm{p}}^2 \sinh \chi) , \label{eq:conventionalH}
\end{eqnarray}
showing the exponential decline for large $\chi$.
 
\section{DIPOLE POTENTIALS}
\label{sec:dipot}

The potentials are defined by the following equations:
\begin{eqnarray}
A_0  & = & h_0 (\chi) P_1 (\theta) \exp (-{\mathrm i} n \eta) ,
\label{eq:Adef0} \\
A_1  & = & h_1 (\chi) P_1 (\theta) \exp (-{\mathrm i} n \eta) ,
\label{eq:Adef1} \\
A_2  & = & h_2 (\chi) N_1 (\theta) \exp (-{\mathrm i} n \eta) ,
\label{eq:Adef2} \\
F_{\mu \nu}  & = & \frac{\partial A_{\nu}}{\partial x^{\mu}} -
\frac{\partial A_{\mu}}{\partial x^{\nu}} \label{eq:Adef3} .
\end{eqnarray}

From (\ref{eq:Adef3}) we derive:
\begin{eqnarray}
h_2 ^{\,\prime} - h_1  & = & f_3 , \label{eq:heq1} \\
h_0 ^{\,\prime} + {\mathrm i} n h_1  & = & f_1 , \label{eq:heq2} \\
h_0 + {\mathrm i} n h_2  & = & f_2 . \label{eq:heq3}
\end{eqnarray}

(\ref{eq:heq1}), (\ref{eq:heq2}) and (\ref{eq:heq3}) are not
independent. If we differentiate (\ref{eq:heq3}), subtract it from
(\ref{eq:heq2}) and use (\ref{eq:dip3}), we obtain (\ref{eq:heq1}).
These three equations therefore do not suffice to define the potentials
uniquely; this is to be expected, because $A_{\mu}$ is only defined up to
a scalar gauge function, $\psi$, so that
$\overline{A}_{\mu} = A_{\mu} + \partial \psi / \partial x^{\mu}$ defines
the same fields as $A_{\mu}$.

To fix the potentials uniquely we need a gauge condition. Although, as we
saw in section \ref{sec:Mink}, there is a real sense in which our final
result is gauge invariant, we still have to specify a gauge condition at
this point to determine how the potentials in our model propagate over
cosmological distances. The natural choice is the Lorenz condition (LC), as
given in \citet{jack2}. This is covariant in form and respects causality;
it also has the advantage that the current density generating the vector
potential is the complete current density, not (as for the Coulomb gauge)
just the transverse part. The curved-space generalization of the LC
is $A^{\mu}_{\, ;\mu} = 0$, which in our metric becomes
\begin{equation}
\frac{\partial}{\partial x^{\mu}}
\left(\sqrt{g}\, g^{\mu \nu} A_{\nu} \right) = 0 .  \label{eq:lrnz}
\end{equation}

This has the disadvantage, in a FRW space, that factors of $R(\eta)$ remain.
This means that the equations describing the propagation of the potentials
will involve the time derivative of $R$, and so will be coupled to the
gravitational field equations. It is certainly possible that Nature works
this way. But to simplify the mathematics we propose here to use the
following modified Lorenz condition (MLC):
\begin{equation}
\frac{\partial}{\partial x^{\mu}}
\left(\sqrt{g}\, g^{\mu \nu} A_{\nu}/R^2 \right) = 0 ,  \label{eq:lrnz2}
\end{equation}
which leads to
\begin{equation}
{\mathrm i}  n \sinh^2 \chi h_0 + \frac{{\mathrm d}  }{{\mathrm d}  \chi}
\left[ \sinh^2 \chi h_1 \right] - 2 h_2 = 0 .
\label{eq:heq4}
\end{equation}

(\ref{eq:lrnz2}) may seem unphysical, because it depends explicitly on
$R(\eta)$, and so seems to be restricted to FRW spaces, rather than being
universally applicable. But it can be given a physical meaning in the
context of conformally invariant theories. The connection is outlined in
appendix \ref{app:conform}.

The potentials are normalized by comparing them with the
expressions for the fields. Together with the MLC, this
determines the potentials uniquely.

\subsection{\label{subsec:A0}Scalar potential}

By combining (\ref{eq:dip1}), (\ref{eq:dip2}), (\ref{eq:heq2}),
(\ref{eq:heq3}) and (\ref{eq:heq4}) we can derive an equation for $h_0$
alone:
\begin{equation}
\frac{{\mathrm d} ^2 h_0}{{\mathrm d}  \chi^2}
+ 2 \coth \chi \frac{{\mathrm d}  h_0}{{\mathrm d}  \chi}
+ \left( n^2 - \frac{2}{\sinh^2 \chi} \right) h_0 = 0 . \label{eq:h0}
\end{equation}

Defining $\alpha = \sqrt{n^2 - 1}$, this becomes the equation for hyperbolic
spherical functions, (\ref{eq:f4}). As before, we choose the solution that
represents outgoing waves. With $u=\tanh (\chi/2)$:
\begin{eqnarray}
A_0 & \equiv & P_1(\theta) h_0 ( \chi ) \exp (-{\mathrm i} n\eta )
\nonumber \\
  & = & C_2 (n) P_1 (\theta) \exp [{\mathrm i} (\alpha \chi - n \eta)] \times
\nonumber \\
  & & \frac{(1 - u^2) \left( 1 - 2{\mathrm i} \alpha u + u^2 \right) }{u^2} ,
\label{eq:h0out1}
\end{eqnarray}
where $C_2 (n)$ is a normalization factor that depends on the driving
function. $C_2 (n)$ is, of course, proportional to the normalization factor
$C_1 (n)$ defined in (\ref{eq:normC1}). For small $\chi$,
(\ref{eq:h0out1}), (\ref{eq:heq3}) and (\ref{eq:f2out}) give:
\begin{equation}
C_2 (n) = \frac{{\mathrm i} C_1 (n)}{4n} . \label{eq:normC2}
\end{equation}

\subsection{\label{subsec:A2}Vector Potential}

The interesting component of the vector potential is the transverse one,
$A_{\theta}$, or in our notation $A_2$. (\ref{eq:heq3}) gives
\begin{equation}
h_2 = \frac{-{\mathrm i}  f_2}{n} + \frac{{\mathrm i}  h_0}{n} . \label{eq:heq2a}
\end{equation}
This can be used to get $h_2$, and from that $A_2$, since 
$f_2$ and $h_0$ are known from (\ref{eq:f2out}) and (\ref{eq:h0out1}).

Equation (\ref{eq:heq2a}) shows that $A_2$ is the sum of two parts, one
involving $\exp ({\mathrm i} n\chi)$, the other involving
$\exp ({\mathrm i} \alpha\chi)$:
\begin{eqnarray}
A_2  & = & \frac{-q V N_1 (\theta) \sin (n \tau)
e ^{-{\mathrm i} n \eta}}
{2 n^3 u^2} ({\cal N} + {\cal A}) , \label{eq:A2basic} \\
{\cal N}  & = & e ^{{\mathrm i} n \chi}
\left[ (1-u^2)^2 - 2{\mathrm i} nu(1+u^2) - 4 n^2 u^2 \right] ,
 \label{eq:calN} \\ 
{\cal A}  & = & e ^{{\mathrm i}  \alpha \chi} (1-u^2)
\left[ -1 + 2{\mathrm i} \alpha u - u^2 \right] . \label{eq:calA}
\end{eqnarray}

\section{PROPAGATION OF THE VECTOR \\
POTENTIAL}
\label{sec:propvec}

The Fourier synthesis of $A_2$ is carried out, as for the magnetic field,
by dividing by $2\pi$ and then integrating over $n$, along a contour chosen
to respect causality. There is no vector potential before the pulse begins,
so the correct contour is a line parallel to the real axis and slightly
above it. The integral must be taken along the whole path A--D in figure
\ref{fig:contour1}, including the semicircle B--C. This ensures that for
$\chi > \eta + \tau$ the contour can be closed in the UHP and the integral
will be zero. There are two regions of interest,
$\eta - \tau < \chi < \eta + \tau$ (the ``main pulse''), and
$\chi < \eta - \tau$ (the ``tail''). 

\begin{figure}
\centering
\includegraphics[scale=0.5]{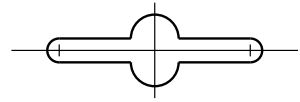}
\caption{\label{fig:contour2} Integration contour for $A_{2}$. The contour
encloses the points $n = -1$ and $n = +1$; it is described in a
clockwise sense. }
\end{figure}

We get a non-zero result only for that part of the integral that involves a
contour that is closed in the LHP. For the part of equation
(\ref{eq:A2basic}) that is proportional to $\exp ({\mathrm i} n \chi)$ we
can shrink this contour to a small circle about the point $n=0$, and we
just have to find the residue there. For the part that is proportional to
$\exp ({\mathrm i}  \alpha \chi)$ we have to remember the branch points at
$n = \pm 1$, so our contour can only be shrunk to the form shown in figure
\ref{fig:contour2}.

\begin{figure}
\centering
\includegraphics[scale=0.5]{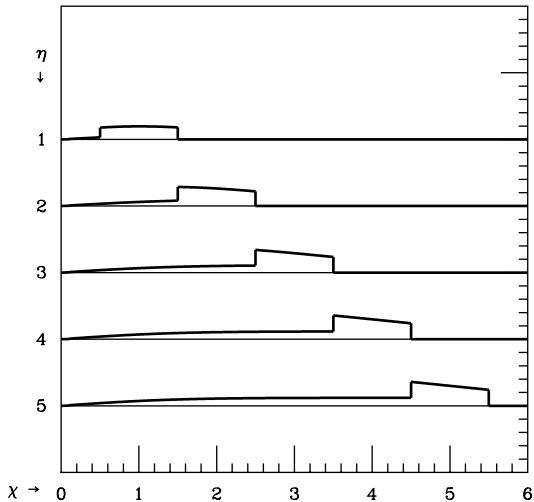}
\caption{\label{fig:ATEPS} Propagation of the transverse component of
the vector potential, $A_2$: snapshots of the pulse and the tail for a
series of values of $\eta$. As with the magnetic field, we set
$\tau = 0.5$ to make the pulse visible on the graph. }
\end{figure}

$A_2$ is the sum of contributions from small circles (or semicircles)
around $n=0$ (the pole terms) and the integrals along the cut from
$n=-1$ to $n=1$. The pole terms have simple analytic expressions, but the
integrals must be evaluated numerically for each ($\eta,\,\chi$) pair.
The integrations are straightforward, and the propagation of $A_2$ is
shown in figure \ref{fig:ATEPS}.

An important difference between figures \ref{fig:HPEPS} and \ref{fig:ATEPS}
is that in the latter the pulse has a non-zero tail for
$\chi < \eta - \tau$. This feature of the propagation of potentials in a
curved space has been noted before (\citet{dew,nar1}; for more recent work
see \citet{hag1}). For dipole
propagation, as here, the tail rises linearly from $\chi = 0$, and
approaches a constant value for large $\chi$. This asymptotic value is
proportional to $\tau$. In ordinary Minkowski space, which corresponds to
the limit $\chi \rightarrow 0$, ${\mathbf A}$ has no tail.

We note that the following simple function, with $a = 0.735$, gives an
adequate fit for $A_2^2$ for large $\eta$:
\begin{equation}
A_2^2 \approx \left( 2 q V \tau \sin \theta \right)^2 
\tanh^2 (a\chi) . \label{eq:A2fit}
\end{equation}

\section{EFFECTIVE THERMAL MASS IN A \\
PLASMA IN FRW SPACE}
\label{sec:FRW}

In section \ref{sec:Mink} we discussed the thermal mass in Minkowski space;
we now have to investigate any changes there may be when we go to a FRW
space. The most important one is the appearance of the tail following the
pulses of ${\mathbf A}$. Since the fields themselves do not have tails,
the vector potential in this region is a pure potential, without
accompanying fields. It can therefore be represented as the gradient of a
scalar function, $\partial {\cal S} / \partial x^{\mu}$.

In the asymptotic region, where both $\eta \gg 1$ and $\chi \gg 1$,
${\cal S}$ has a simple form. Both $h_0$ and $h_1$ tend to zero
exponentially as $\chi \rightarrow \infty$. $h_2$ is given by the contour
integral of the previous section, and in the asymptotic region the integral
along the cut is negligible compared to the pole term. So we get
$A_2 \approx 2qV \tau N_1 (\theta)$, and
${\cal S} \approx 2qV \tau P_1 (\theta)$.

If the potential from source particle $i$ is denoted by $A_{2i}$, then 
the total from all the particles in any one configuration is
${\cal S} = \sum_i {\cal S}_i$. We might think that we could use ${\cal S}$
as a gauge function and remove the effect of $A_2$ in the tail entirely.
But this is not so once we introduce a density matrix and sum over all
configurations of the distant source particles. Each configuration now
produces its own gauge function, ${\cal S}_{\mu} = \sum_i {\cal S}_{\mu i}$.
These functions are all distinct, and there is no single gauge function that
can be used to eliminate the effect of $A_2$ in the tail.

This is not to say gauge invariance is violated; as discussed in section
\ref{sec:Mink}, we can introduce an arbitrary overall gauge function
$\psi (x^{\mu})$. But this is quite different from the separate functions
${\cal S}_{\mu i}$, each of which is determined by the position and
orientation of the corresponding source dipole.

\section{MASS GENERATION: FIRST \\
CALCULATION}
\label{sec:mass_1}

We will show here that $\langle A^{\mu} A_{\mu} \rangle$ is composed of a
thermal part, $\langle A^{\mu} A_{\mu} \rangle_{\mathrm{th}}$, which will
be of order $T^2$, and a non-thermal part,
$\langle A^{\mu} A_{\mu} \rangle_{\mathrm{nt}}$.
We are concerned with the ratio
\begin{equation}
p = \langle A^{\mu} A_{\mu} \rangle_{\mathrm{nt}}/
\langle A^{\mu} A_{\mu} \rangle_{\mathrm{th}} . \label{eq:p_ratio}
\end{equation}
and its variation with time.

Moving particles in the distant plasma will generate pulses of $A_{\mu}$,
each of which consists of a ``main pulse'' and a ``tail''. The main pulse
contains electric and magnetic fields, which are distinguished from the
thermal background by their spectrum. The associated vector potential will
contribute to $\langle A^{\mu} A_{\mu} \rangle_{\mathrm{nt}}$. We will not
investigate this in detail, because, as we will see later, it is the
potentials in the tail, not the main pulse, that are responsible for the
increase of $p$ with time. $A_{\mu}$ in the tail is a pure gauge potential
that generates no fields, but it, too, will contribute to
$\langle A^{\mu} A_{\mu} \rangle_{\mathrm{nt}}$.

To study the development of $\langle A^{\mu} A_{\mu} \rangle_{\mathrm{nt}}$
we need a model of how the Universe came into being. We will suppose it
originated as a bubble in a metastable vacuum \citep{buch1}, and for
simplicity assume that all fields at temperature $T_{\mathrm{max}}$ 
appeared without appreciable delay at the surface of the bubble.
The evolution of $\langle A^{\mu} A_{\mu} \rangle_{\mathrm{nt}}$ can be
visualized as follows. As each pulse passes the observation point it leaves
a memory in the form of the tail. These tails will not cancel but will add
according to the theory of random flights \citep{hugh1}. The scalar
potential, of course, will remain close to zero; it is only the vector
potential that accumulates these additions. Consequently, from now on we
will write $\langle A^{\mu} A_{\mu} \rangle_{\mathrm{th}} =
\langle {\mathbf A}^2 \rangle_{\mathrm{th}}$, and similarly for the
non-thermal part.

In (\ref{eq:p_ratio}), both the numerator and the denominator will decrease
as the temperature falls, but the numerator falls more slowly, so the ratio
will gradually increase from zero. In some cosmological models, as we will
see in section \ref{sec:mannheim_model}, we would like to know when $p$
becomes of order unity, so we will estimate this here.

\begin{figure}
\centering
\includegraphics[scale=0.5]{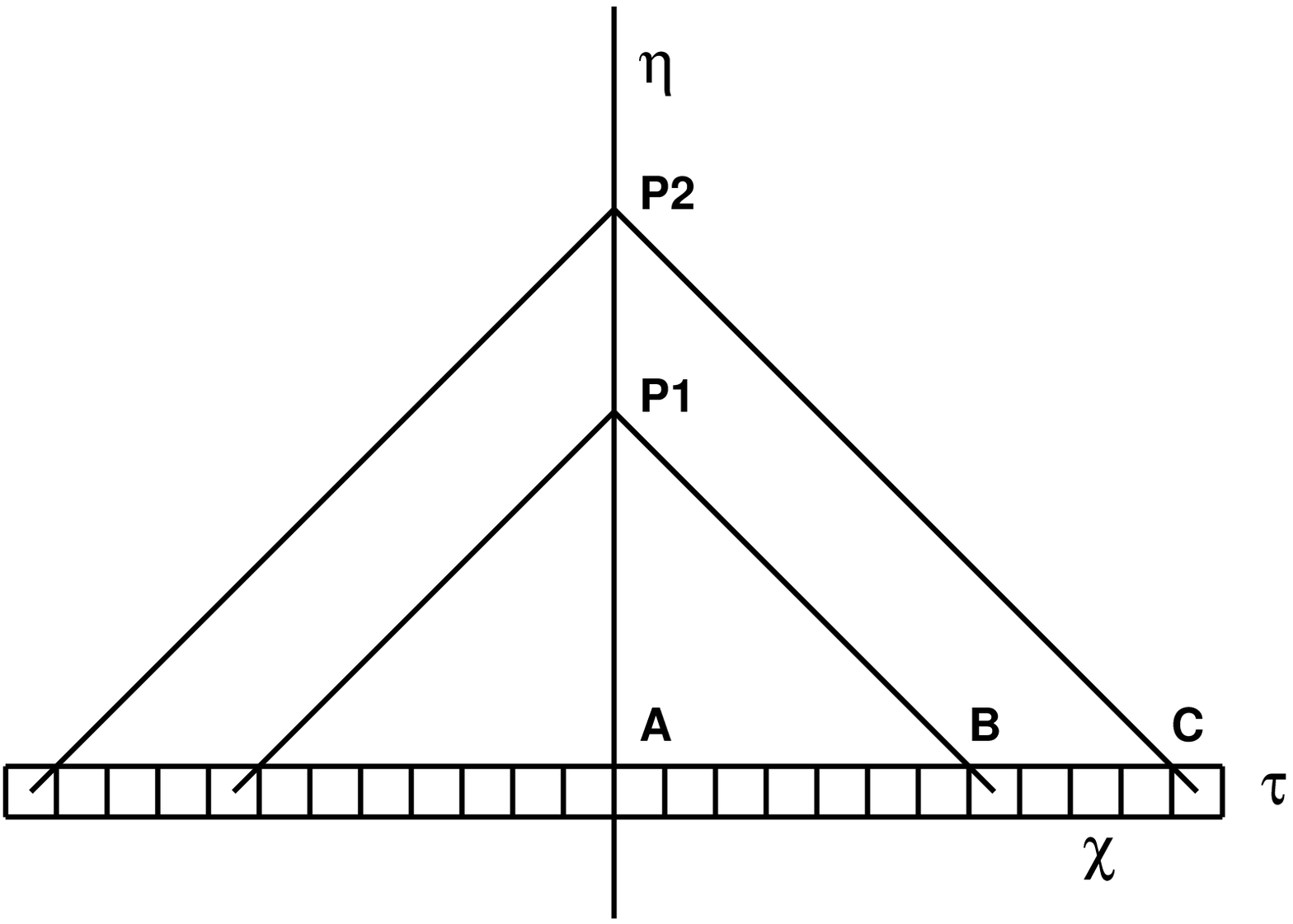}
\caption{\label{fig:build} Buildup of
$\langle {\mathbf A}^2 \rangle_{\mathrm{nt}}$. Each box represents a
spherical shell of radius $\chi$, thickness $\mathrm{d} \chi$, duration
$\tau$. For clarity, only one time slice is shown; the plasma actually
fills the whole volume from $\eta = 0$ to the observation time.  }
\end{figure}

It is convenient at this point to change coordinates so the observation
point is at $\eta = \eta_p$, $\chi = 0$, and a general source point is at
$\eta$, $\chi$. Our formulae involve differences in $\eta$ and $\chi$, which
are unchanged by this shift of origin. We will consider $\eta$ and $t$ as
starting from zero at the time of minimum radius, $R_{\mathrm{min}}$, and
maximum temperature, $T_{\mathrm{max}}$. We can picture the buildup of
$\langle {\mathbf A}^2 \rangle_{\mathrm{nt}}$
with the help of figure \ref{fig:build}, where
the coordinates are $\eta$ (upwards) and $\chi$. The row of boxes at the
bottom of the diagram is at $\eta = 0$; this represents the surface of the
bubble. Each box has thickness ${\mathrm d}  \chi$ and duration $\tau$. Two
observation points are shown, ${\mathrm{P}}_1$ and ${\mathrm{P}}_2$, at
times $\eta_1$ and $\eta_2$. The lines ${\mathrm{P}}_1$--B and
${\mathrm{P}}_2$--C represent the past light cones. All plasma particles
within these past light cones contribute to
$\langle {\mathbf A}^2 \rangle_{\mathrm{nt}}$. The contributions add
incoherently over times longer than the collision time, $\tau$, so we can
imagine the whole diagram divided into time slices of duration $\tau$, just
like the one shown for $\eta = 0$. A box at source time
$\eta_{\mathrm{s}}$ represents a spherical shell of volume
$4\pi R_{\mathrm{s}}^3 \sinh^2 \chi \, {\mathrm d}  \chi$.

The number density of particles in this shell, for a thermal distribution, is
\begin{equation}
\nu_{\mathrm{s}} = 0.24 T_{\mathrm{s}}^3 , \label{eq:numdensity}
\end{equation}
and the number of particles in the shell is then
\begin{eqnarray}
{\mathrm d} N_{\mathrm{s}} & = & 4 \pi (0.24 T_{\mathrm{s}}^3)
R_{\mathrm{s}}^3 \sinh^2 \chi \, {\mathrm d}  \chi \nonumber \\
  & \equiv & 4 \pi (0.24) A_{\mathrm{M}}^{3/4} \sinh^2 \chi \,
{\mathrm d}  \chi .  \label{eq:numshell}
\end{eqnarray}
Here we have introduced the quantity $A_{\mathrm{M}} = R^4 T^4$, which (at
our current level of calculation) will be constant during the expansion,
and therefore does not need another subscript, `s' or `p'.

Multiplying (\ref{eq:numshell}) and (\ref{eq:A2fit}) we get the contribution
to $A_2^2$ from the particles in the shell. We convert this to
$g^{\mu\nu} A_{\mu} A_{\nu}$ by multiplying by $g^{22}$:
\begin{equation}
4\pi (0.24) A_{\mathrm{M}}^{3/4} \sinh^2 \chi
\, {\mathrm d}  \chi \left(2q \tau \right)^2
\langle V^2 \rangle \langle \sin^2 \theta \rangle 
\frac{\tanh^2 (a \chi)}{R_{\mathrm{p}}^2 \sinh^2 \chi} .
\label{exp:A2shell}
\end{equation}

The average of $\sin^2 \theta$ over a sphere gives $2/3$. For the average of
$V^2$ we reason as follows: if the effective $m^2$ of the particles were
purely thermal, we could write $V^2 = 1/2$. But the effective mass
actually increases with time as the ratio $p$ from (\ref{eq:p_ratio})
increases from zero towards one. So a better estimate of $V^2$ is
$1/(2+p)$.

$\eta = 0$ at the surface of the bubble, and $\eta = \eta_{\mathrm{p}}$ at
the observation point. For some time slice at time $\eta$ between these two
limits,  we get the contribution of all spherical shells
to $\langle {\mathbf A}^2 \rangle_{\mathrm{nt}}$ by integrating from
$\chi = 0$ to the light cone at $\chi = \eta_{\mathrm{p}} - \eta$:
\begin{equation}
\frac{8\pi (0.24)}{3} A_{\mathrm{M}}^{3/4} \left(\frac{1}{2+p}\right)
\left(2q \tau \right)^2 \int_0^{\eta_{\mathrm{p}} - \eta} \!\!
{\mathrm d}  \chi \frac{\tanh^2 (a \chi)}{R_{\mathrm{p}}^2 } .
\label{exp:A2slice}
\end{equation}

We now have to sum over all time slices of duration $\tau$. The sum can be
converted to an integral by multiplying by ${\mathrm d}  \eta/\tau$:
\begin{eqnarray}
\frac{32\pi (0.24)}{3} A_{\mathrm{M}}^{3/4} q^2
\int_0^{\eta_{\mathrm{p}}} \!\! {\mathrm d}  \eta
\left(\frac{1}{2+p}\right) \tau \times \nonumber \\
\int_0^{\eta_{\mathrm{p}} - \eta}
\!\! {\mathrm d}  \chi \frac{\tanh^2 (a \chi)}{R_{\mathrm{p}}^2 } .
\label{exp:A2both}
\end{eqnarray}

In section \ref{sec:source} we defined $\tau = \tau_t / R_{\mathrm{s}} (t)$.
For $\tau_t$ we will simply use equation 3.24 from \citet{mont2}. The
factor $U^3$ in the numerator will be of order $(2+p)^{-3/2}$; the quantity
in the bracket in the denominator is of order unity and will be omitted. We
then have, in our notation,
$\tau_t = m^2(2+p)^{-3/2}/(8\pi \nu_{\mathrm{s}} q^4 \ln \Lambda)$.
For $m^2$ we will use the effective mass at time $t$. The purely thermal
value is $q^2 T_{\mathrm{s}}^2$, but we should multiply this by $(1+p)$ to
include the non-thermal part also. We have
$\nu_{\mathrm{s}} = 0.24 T_{\mathrm{s}}^3$, so
\begin{eqnarray}
\tau & = & \frac{1+p}
{8 \pi \ln \Lambda (0.24) T_{\mathrm{s}} q^2 R_{\mathrm{s}} (2+p)^{3/2} }
\nonumber \\
  & = & \frac{1+p}
{8 \pi \ln \Lambda (0.24) q^2 A_{\mathrm{M}}^{1/4} (2+p)^{3/2} } .
\label{eq:tau_est}
\end{eqnarray}

We can now get an expression for $p$ by using (\ref{eq:tau_est}) in
(\ref{exp:A2both}) and dividing by $T_{\mathrm{p}}^2$:
\begin{eqnarray}
p & = & \frac{4}{3 \ln \Lambda}
A_{\mathrm{M}}^{3/4} q^2 \int_0^{\eta_{\mathrm{p}}}
\!\! {\mathrm d}  \eta
\left(\frac{1+p}{ q^2 A_{\mathrm{M}}^{1/4} (2+p)^{5/2} } \right) 
\times \nonumber \\
  & & \int_0^{\eta_{\mathrm{p}} - \eta} \!\! {\mathrm d}  \chi
\frac{\tanh^2 (a \chi)}{R_{\mathrm{p}}^2 T_{\mathrm{p}}^2} \nonumber \\
 & = & \frac{4}{3 \ln \Lambda}
\int_0^{\eta_{\mathrm{p}}} \!\! {\mathrm d}  \eta 
\frac{1+p}{(2+p)^{5/2}}
\int_0^{\eta_{\mathrm{p}} - \eta} \!\! {\mathrm d}  \chi \tanh^2 (a \chi) 
\nonumber \\
 & = & \frac{4}{3 \ln \Lambda} \int_0^{\eta_{\mathrm{p}}} \!\!
{\mathrm d}  \eta \frac{1+p}{(2+p)^{5/2}} \times \nonumber \\
  & & \left\{\eta_p - \eta - \tanh [a(\eta_p - \eta)]/a \right\} .
\label{eq:p_exp_1}
\end{eqnarray}

This integral equation is surely not correct in detail, but it does provide
a general picture of the buildup of $p$ from $p=0$ at $\eta_p=0$.
We seek $\eta_{\mathrm{cw}}$, the value of $\eta_p$ for which $p=1$. Use
$a = 0.735$, as before, and for definiteness set $\ln \Lambda = 10$;
numerical solution then gives $\eta_{\mathrm{cw}} \approx 11$.

A remarkable feature of (\ref{eq:p_exp_1}) is that not only has
$A_{\mathrm{M}}$ disappeared, but also the coupling constant, $q$. The
long-range character of the Coulomb interaction, however, is still apparent
in $\ln \Lambda$. An equation of this sort might reasonably have been
expected to yield a value of $\eta_{\mathrm{cw}}$ that was either extremely
small, so the CW transition takes place almost immediately after the
formation of the bubble, or extremely large, so the transition never takes
place at all.  Instead we get a value of $\eta_{\mathrm{cw}}$ that is within
one or two orders of magnitude of unity.

\section{INTRODUCTION OF CONDUCTIVITY}
\label{sec:introcond}

Up to this point we have been mainly concerned with potentials in the tail.
But we have now to look more closely at the electric and magnetic fields in
the main pulse for dipole propagation. These combine to give a Poynting
vector directed outwards. In flat space, this vector falls off like
$1/\chi^4$, which tends to zero even when integrated over the whole sphere.
When we go to curved space, however, the Poynting vector tends to
$(q V \sin \theta / \sinh \chi )^2$. The surface area of a sphere is 
$4 \pi R^2 \sinh^2 \chi$, so the integral of the Poynting vector tends to a
constant non-zero value. One effect of curvature is to continuously increase
the energy of the plasma.

It seems unlikely this extra energy will remain in the form of pulses like
those of figure \ref{fig:HPEPS}. We know such pulses propagate unchanged in
a flat space, but when they have traveled cosmological distances, so the
curvature becomes noticeable, we should expect a gradual thermalization.
We will model this thermalization by including a small, constant
conductivity, $\sigma$, in our equations. Note that $\sigma$ is not directly
related to the normal conductivity of the plasma, which (in our cosmological
units) would be very high. $\sigma$ is just a device to represent the slow
thermalization, and will have a value of order unity. For technical reasons,
which we explain below, we choose $\sigma = 1/(2\pi)$.

There are two reasons why we have to investigate the effect of $\sigma$:
\begin{enumerate}
	\item The thermalization of the pulses raises the temperature of the
	plasma, and so tends to reduce the value of the ratio $p$.
	\item The slow decline in the height of the pulses will probably reduce
	the value of $A$ in the tail, and this will also reduce the value of $p$.
\end{enumerate}

The gradual transfer of energy to the plasma implies that the product 
$R(\eta) T(\eta)$ will not remain constant, as in a simple expansion, but
will slowly increase. In this respect $\sigma$ simulates inflation, but 
on a much longer time scale.

The equations involved here are analogous to those used previously, but
are considerably more complicated, so we just give the main results.

\section{INCLUSION OF CONDUCTIVITY: \\
FIELDS}
\label{sec:incl_conduct}

The wave number, $k$, the permittivity, $\epsilon (n)$, and the
conductivity, $\sigma$, satisfy the dispersion relation
\begin{equation}
k^2 = n^2 \epsilon(n) = n(n + 4\pi {\mathrm i}  \sigma) .
\label{eq:dispersion}
\end{equation}

By analogy with Maxwell's equations in flat space, the equations
(\ref{eq:dip1}), (\ref{eq:dip2}) and (\ref{eq:dip3}) for our
dipole fields become:
\begin{eqnarray}
{\mathrm i} n \epsilon \sinh^2 \chi f_1 - 2 f_3  & = & 0 , \label{eq:dip1perm} \\
{\mathrm i} n \epsilon f_2 - f_3^{\,\prime}  & = & 0 , \label{eq:dip2perm} \\
f_1 - f_2^{\,\prime} + {\mathrm i} n f_3  & = & 0 . \label{eq:dip3perm} 
\end{eqnarray}

\subsection{\label{subsec:conduct_H}Magnetic Field}

From these we obtain an equation for $f_3$ alone:
\begin{equation}
\frac{{\mathrm d} ^2 f_3}{{\mathrm d}  \chi^2} +
\left(n^2 \epsilon - \frac{2}{\sinh^2 \chi} \right) f_3 = 0 .
\label{eq:f3perm}
\end{equation}

Writing $k^2$ for $n^2 \epsilon$ this equation takes a familiar form; $f_3$
can be derived from the empty-space formula simply by writing $k$ for $n$
everywhere, except, of course, for the normalization function,
which we denote now by $C_3 (n,\sigma)$:
\begin{equation}
f_3 = \frac{C_3 (n,\sigma) \exp ({\mathrm i} k\chi)}{2 u}
\left(1 - 2 {\mathrm i} ku + u^2 \right) . \label{eq:f3out2perm}
\end{equation}

$C_3 (n,\sigma)$ is most easily determined in flat space; this is sufficient
since we only have to consider small distances. The calculation is done in
appendix \ref{app:normalize}. We find:
\begin{equation}
C_3 (n,\sigma) = \frac{-6qV (n + 4\pi {\mathrm i}  \sigma) \sin (n \tau)}
{n (3n + 8 \pi {\mathrm i}  \sigma)} .
\label{eq:normC3perm}
\end{equation}

\subsection{\label{subsec:propf12perm}Propagation of the Magnetic Field}

The Fourier transform of the magnetic field becomes
\begin{eqnarray}
F_{12} & = & \frac{-3qV (n + 4\pi {\mathrm i}  \sigma ) N_1 (\theta)
\sin (n\tau) \exp ({\mathrm i} k\chi - {\mathrm i} n\eta)}
{(3n + 8 \pi {\mathrm i}  \sigma ) n u} \times \nonumber \\
  & & \left(1 - 2 {\mathrm i} ku + u^2 \right) . \label{eq:F12perm}
\end{eqnarray}

\begin{figure}
\centering
\includegraphics[scale=0.5]{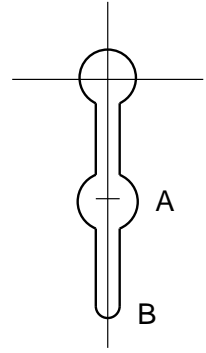}
\caption{\label{fig:CONTOUR2EPS}Contour for computing the magnetic field,
$F_{12}$, when conductivity is included. The points on the negative
imaginary axis labeled $A$ and $B$ are at $n = -8\pi {\mathrm i}  \sigma / 3$
and $n = -4 \pi {\mathrm i}  \sigma$, respectively. The contour is traversed
in a clockwise sense. }
\end{figure}

We can integrate around the pole at $n=0$ in the same way as before, except
that we have to respect the branch points of $k$, at $n=0$ and
$n=-4\pi {\mathrm i}  \sigma$. We have also to take account of the pole at
$n = -8 \pi {\mathrm i}  \sigma / 3$. A suitable contour is shown in figure
\ref{fig:CONTOUR2EPS}.

\begin{figure}
\centering
\includegraphics[scale=0.5]{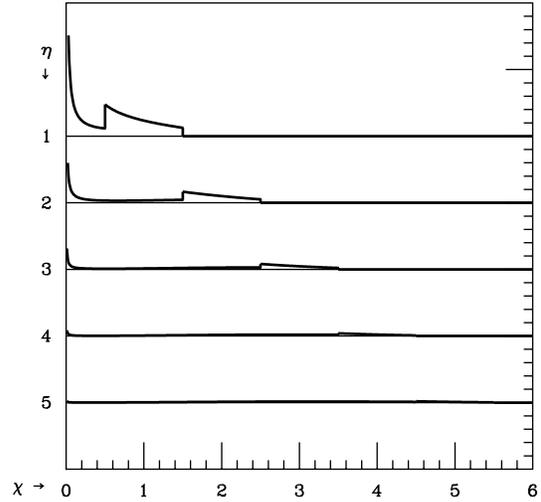}
\caption{\label{fig:HP2EPS}Propagation of the magnetic field, $F_{12}$,
in the presence of conductivity, $\sigma = 1/2\pi$. }
\end{figure}

The integrals are straightforward, and the resulting propagation of
$F_{12}$ is shown in figure \ref{fig:HP2EPS}.

\section{INCLUSION OF CONDUCTIVITY: \\
POTENTIALS}
\label{sec:conduct_A}

Here we follow the prescription of Landau and Lifshitz \citep{landau2}:
use the same equations relating potentials and fields as in empty space,
i.e. (\ref{eq:heq1}), (\ref{eq:heq2}) and (\ref{eq:heq3}), but modify the
Lorenz condition by including the permittivity in the $h_0$ term:
\begin{equation}
{\mathrm i}  n \epsilon \sinh^2 \chi h_0 + \frac{{\mathrm d} }{{\mathrm d} \chi}
\left[\sinh^2 \chi h_1 \right] - 2 h_2 = 0 .  \label{eq:heq4perm}
\end{equation}

\subsection{\label{subsec:conduct_h0}Scalar Potential}

From these equations, as before, we can derive an equation for $h_0$ alone:
\begin{equation}
\frac{{\mathrm d} ^2 h_0}{{\mathrm d}  \chi^2}
+ 2 \coth \chi \frac{{\mathrm d}  h_0}{{\mathrm d}  \chi}
+ \left(n^2 \epsilon - \frac{2}{\sinh^2 \chi} \right) h_0 = 0 .
\label{eq:h0perm}
\end{equation}
This is the same equation as we obtained for empty space, except that, just
as for the magnetic field, in place of $n^2$ we must write
$k^2 \equiv n^2 \epsilon$. (\ref{eq:h0out1}) applies as before, provided we
write $\alpha = \sqrt{k^2 - 1}$, and use a normalization function $C_4
(n,\sigma )$ in place of $C_2 (n)$: 
\begin{equation}
C_4 (n,\sigma ) = \frac{{\mathrm i} C_3 (n,\sigma )}
{4(n + 4 \pi {\mathrm i}  \sigma )} .  \label{eq:normC4perm}
\end{equation}

\subsection{\label{subsec:conduct_h2}Vector Potential}

Defining $n_p = -8 \pi {\mathrm i} \sigma /3$, we find:
\begin{eqnarray}
A_2 & = & \frac{-q V N_1 (\theta) \sin (n \tau) e ^{-{\mathrm i} n \eta}}
{2 (n - n_p ) n^2 u^2}
\left( {\cal K_{\mathrm{\sigma}} } +{\cal A_{\mathrm{\sigma}} } \right) .
\label{eq:A2basicperm} \\
{\cal K_{\mathrm{\sigma}}}  & = & e ^{{\mathrm i}  k \chi } 
\left[(1-u^2)^2 - 2{\mathrm i} ku (1+u^2) - 4k^2u^2 \right] ,  
\label{eq:calKperm} \\ 
{\cal A_{\mathrm{\sigma}}}  & = & e ^{{\mathrm i}  \alpha \chi } (1-u^2)
\left( -1 + 2{\mathrm i} \alpha u - u^2 \right) . \label{eq:calAperm}
\end{eqnarray}

We note that going from (\ref{eq:A2basic}) to (\ref{eq:A2basicperm})
takes three simple steps:
\begin{enumerate}
	\item Rename ${\cal N}$ and ${\cal A}$; call them ${\cal K}_{\sigma}$
	and ${\cal A}_{\sigma}$, respectively.
	\item In the final parenthesis of (\ref{eq:A2basic}), substitute $k$
	for $n$ in both ${\cal K}_{\sigma}$ and ${\cal A_{\sigma}}$; this
	includes redefining $\alpha$ to be $\sqrt{k^2 - 1}$ rather than
	$\sqrt{n^2 - 1}$.
	\item In the prefactor of (\ref{eq:A2basic}), change $n^3$ in the
	denominator to $n^2 (n - n_p)$.
\end{enumerate}

(\ref{eq:A2basicperm}) becomes (\ref{eq:A2basic}) in the limit
$\sigma \rightarrow 0$, as it should.

\subsection{\label{subsec:conduct_propA2}Propagation of the Vector
Potential}

For the Fourier synthesis we need, in general, to use a more complicated
contour than the one shown in figure \ref{fig:contour2}, because we
have to take account of the branch points of both $k$ and $\alpha$. With our
choice of $\sigma = 1/2\pi$, however, the branch points of $\alpha$ coalesce
into a single point at $n=-{\mathrm i} $, and the contour of figure
\ref{fig:contour2} is adequate.

\begin{figure}
\centering
\includegraphics[scale=0.5]{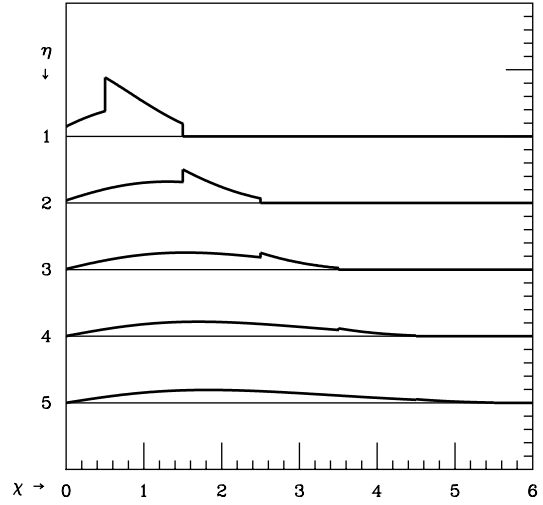}
\caption{Propagation of the vector potential, $A_2$, in the presence of
conductivity, $\sigma = 1/2\pi$. }
\label{fig:AT2EPS}
\end{figure}

The result is shown in figure \ref{fig:AT2EPS}.
 
\subsection{\label{subsec:A2asymptotic}Asymptotic Form of the
Vector Potential}

We can show easily that in the limit $\eta \rightarrow \infty$,
$A_2$ satisfies a diffusion equation. The main terms come from the integral
down the cut, and in this limit the integrand is concentrated close to
$n=0$. Note that this integral involves only ${\cal K_{\sigma}}$, the
``field term''; the ${\cal A_{\sigma}}$ (``scalar potential'') term, being
even in $n$, does not contribute. From this integral we obtain a simple
analytic expression for the asymptotic form:
\begin{eqnarray}
A_2  & \approx & \frac{3 q V \tau \sin \theta }{16 \pi}
F( \sigma, \chi, \eta ) , \label{eq:A2asymptotic_2} \\
F( \sigma, \chi, \eta )  & = & \left\{ \frac{(1-u^2)^2}{ \sigma u^2 }
{\mathrm{erf}} \left(\chi \sqrt{\frac{\pi \sigma}{\eta}} \right)
\right.  \nonumber \\
  & & \hspace{-7em} \left. {} - \frac{4 }{\sigma^{1/2} \eta^{3/2} u}
\left[ (1+u^2)\eta + 4\pi \sigma u \chi \right] 
\exp \left(\frac{-\pi \sigma \chi^2 }{\eta} \right) \right\} .
\label{eq:Fasymptotic_2}
\end{eqnarray}

When $\chi$ is also large, of order $\sqrt{\eta}$, the term involving
the error function is negligible compared to the others.

\begin{figure}
\centering
\includegraphics[scale=0.5]{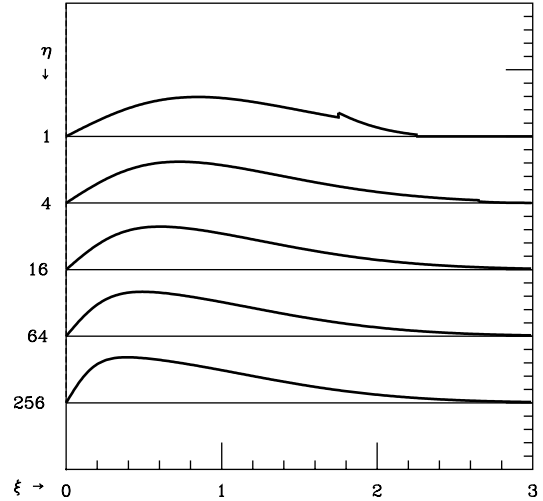}
\caption{$\sqrt{\eta} A_2$ plotted against $\xi = \chi /\sqrt{\eta}$,
to demonstrate the asymptotic form. In this graph we use the asymptotic
form (\ref{eq:A2asymptotic_2}) for the tail. }
\label{fig:AT3EPS}
\end{figure}

In figure \ref{fig:AT3EPS} we plot $\sqrt{\eta} A_2$ against
$\xi = \chi/\sqrt{\eta}$, with $A_2$ computed using the asymptotic formula.
We see $\sqrt{\eta} A_2$ tending to a constant form; with our choice of
abscissa, this is a Gaussian for large $\chi$ and $\eta$. For $\eta \ge 4$,
this graph is indistinguishable from the one using integration around the
contour.

\section{THE ENERGY-TRANSFER EQUATION}
\label{sec:energy_transfer}

We will now investigate the effect of a non-zero conductivity on the
temperature of the plasma. We start from \citet{wein2}, equation 4.7.9:
\begin{equation}
T^{\mu\nu}_{\;\;\; ;\mu} = \frac{1}{\sqrt{g}}
\frac{\partial}{\partial x^{\mu}} \left( \sqrt{g} T^{\mu\nu} \right) +
\Gamma^{\nu}_{\mu\lambda} T^{\mu\lambda} . \label{eq:wein4_7_9}
\end{equation}
This expression is zero for any value of $\nu$, but the most important one
for us is $\nu = 0$. We treat the plasma as a perfect fluid at rest in
FRW coordinates $\eta$, $\chi$, $\theta$, $\phi$, so that the 4-velocity
$\left( U^0 , U^i \right)$ has $U^i = 0$, $i=1,2,3$. Since
$g_{\mu \nu } U^{\mu} U^{\nu} = -1$ and $g_{00} =  R^2 (\eta)$, 
$U^0 = 1/R$.

Multiply (\ref{eq:wein4_7_9}) by $\sqrt{g}\,{\mathrm d} ^4 x\,U_{\nu}$,
sum over $\nu$, and set the resulting expression equal to zero; we get
\begin{equation}
\sqrt{g}\, {\mathrm d}^4 x\, U_0 T^{\mu 0}_{\;\;\; ;\mu} = 0 .
\label{eq:econ1}
\end{equation}
The left-hand side of this equation is a coordinate scalar with
$\sqrt{g}\, {\mathrm d}^4 x$ an element of proper volume. Expanding
(\ref{eq:econ1}):
\begin{eqnarray}
\sqrt{g}\, {\mathrm d} ^4 x\, U_0 \left[ \frac{1}{\sqrt{g}}
\frac{\partial}{\partial x^0 } \left( \sqrt{g}\, T^{00} \right) 
+ \frac{1}{\sqrt{g}} \frac{\partial}{\partial x^i } 
\left( \sqrt{g}\, T^{i 0} \right) + \right. \nonumber \\
\left. \Gamma^{0}_{ij} T^{ij}
+ \Gamma^{0}_{00} T^{00} \right] = 0 . \label{eq:econ2}
\end{eqnarray}

From this we can derive an integrodifferential equation for  $T^{00}$.
The resulting equation is most simply written in terms of
$A_{\mathrm{M}} (\eta) = R^4 (\eta) T^4 (\eta) \,\mathrm{(eV)^4 \cdot m^4}$.
A similar quantity was introduced in section \ref{sec:mass_1}, but there it
could be treated simply as a constant. Here we calculate its variation with
time:
\begin{eqnarray}
{}  \dot{A}_M (\eta ) -
P_4 \int_0^{\eta} {\mathrm d}  \eta^{\,\prime}
A_{M}^{3/4} (\eta^{\,\prime})
\frac{\sigma (\eta - \eta^{\,\prime})}{2 + p(\eta )} \times & & \nonumber \\
\exp \left[ -2 \pi \sigma (\eta - \eta^{\,\prime}) \right] & = & 0 ,
\label{eq:econ6} \\
{}  P_4 = 3.97 \times 10^{-7}\,\mathrm{eV \cdot m} & & .
\label{eq:P4def} 
\end{eqnarray}

Note that because of the appearance of the function $p(\eta )$, this equation
by itself is not sufficient to calculate $A_{\mathrm{M}}$. We obtain an
independent equation connecting these two functions in the next section.

\section{MASS GENERATION: SECOND \\
CALCULATION}
\label{sec:mass_2}
 
We will now obtain an equation for the ratio $p (\eta)$ analogous to
(\ref{eq:p_exp_1}), but based on the asymptotic form
(\ref{eq:A2asymptotic_2}) for the tail of the vector potential. In our
previous calculation, where the tail of $A_2$ reached a constant asymptotic
value, we could be sure that $p$ would rise monotonically; the only question
was whether there would be factors of $A_{\mathrm{M}}$ that would make the
increase far too fast or far too slow. (This is just what happens, for
example, if we use quadrupole rather than dipole potentials.) In this second
calculation, we can anticipate that there will not be any unwanted factors
of $A_{\mathrm{M}}$, but only a detailed calculation will show whether $p$
continues to rise with $\eta$.

The analog of (\ref{eq:numshell}) is
\begin{equation}
{\mathrm d} N_{\mathrm{s}} = 4 \pi (0.24)
A_{\mathrm{M}}^{3/4} (\eta_{\mathrm{s}} )
\sinh^2 \chi \, {\mathrm d}  \chi .  \label{eq:numshell_2}
\end{equation}

To get the contribution to $A_2^2$ at $\eta_{\mathrm{p}}$ from the
particles in the shell, we multiply (\ref{eq:numshell_2}) by
$A_2^2 [ \sigma, \chi, (\eta_{\mathrm{p}} - \eta_{\mathrm{s}} )]$,
with $A_2 ( \sigma, \chi, \eta )$ given by (\ref{eq:A2asymptotic_2}).

Proceeding as before we get the analog of (\ref{exp:A2both}):
\begin{eqnarray}
\frac{2\pi (0.24) q^2 }{3 R^2 (\eta_{\mathrm{p}}) }
\int_0^{\eta_{\mathrm{p}}} {\mathrm d}  \eta_{\mathrm{s}}
\left(\frac{ A_{\mathrm{M}}^{3/4} (\eta_{\mathrm{s}}) }
{2+p (\eta_{\mathrm{s}})}\right) \tau  \times & & \nonumber \\
\int_0^{\eta_{\mathrm{p}} - \eta_{\mathrm{s}}} \!\! {\mathrm d}  \chi
F^2 \left[\sigma,\chi,(\eta_{\mathrm{p}} - \eta_{\mathrm{s}})\right]  & & ,
 \label{exp:A2both_2}
\end{eqnarray}
with $F_0$ given by (\ref{eq:Fasymptotic_2}).

The collision time, $\tau$, will be given by
\begin{equation}
\tau = \frac{(1+p)}{8 \pi \ln \Lambda (0.24) q^2
A_{\mathrm{M}}^{1/4} (\eta_{\mathrm{s}}) (2+p)^{3/2} } .
\label{eq:tau_est_2}
\end{equation}

We can now get an expression for $p$ by using (\ref{eq:tau_est_2}) in
(\ref{exp:A2both_2}) and dividing by $T^2 (\eta_{\mathrm{p}})$:
\begin{eqnarray}
p (\eta_{\mathrm{p}})  & = & 
\frac{3 }{256 \pi^2 \ln \Lambda A_{\mathrm{M}}^{1/2}(\eta_{\mathrm{p}}) }
\int_0^{\eta_{\mathrm{p}}} \!\! {\mathrm d}  \eta_{\mathrm{s}} \,
\times  \nonumber \\
 & & \hspace{-5em} \left( \frac{ \left[1+p (\eta_{\mathrm{s}})\right]
A_{\mathrm{M}}^{1/2} (\eta_{\mathrm{s}}) }
{\left[2+p (\eta_{\mathrm{s}})\right]^{5/2} }\right) 
 \int_0^{\eta_{\mathrm{p}} - \eta_{\mathrm{s}}}
\!\! {\mathrm d}  \chi F^2 \left[\sigma,\chi,
(\eta_{\mathrm{p}} - \eta_{\mathrm{s}})\right]  , \label{eq:pinteqn_2}
\end{eqnarray}
with $F$ given by (\ref{eq:Fasymptotic_2}).

\subsection{\label{subsec:pandAM}Calculation of $\mathbf{p}$
and $\mathbf{A}_{\mathrm{M}}$}

As before, we set $\ln \Lambda = 10$ for definiteness. Simultaneous
numerical solution of (\ref{eq:econ6}) and (\ref{eq:pinteqn_2}) is
straightforward, but we need a starting value for
$A_{\mathrm{M}}$. We choose
$A_{\mathrm{M}} = 2.5 \times 10^{74}\, \mathrm{eV^4 \cdot m^4}$. This is
approximately equal to the observed value today. $p$ becomes equal to unity
around $\eta_{\mathrm{cw}} = 480$.

This value of $\eta_{\mathrm{cw}}$ is significantly larger than the
previous value of $11$ when $\sigma = 0$, but is still within a few
orders of magnitude of unity. The most important feature of this calculation
is that it shows $p(\eta)$ does continue to rise, even when $\sigma \ne 0$.

\section{APPLICATION TO THE MANNHEIM \\
MODEL}
\label{sec:mannheim_model}

In a series of papers (see \citet{mann6} and references given there),
Mannheim has developed a cosmology based on conformal gravity. The
underlying geometry is a FRW space with negative curvature. The expansion
parameter, $R(\eta)$, has a minimum, non-zero value; we will assume this
represents the initial size of the bubble.

A distinguishing aspect of this theory is that it contains no intrinsic
mass scale, unlike conventional cosmology where the Planck mass provides a
fundamental scale. It is therefore natural to ask whether, in Mannheim's
model, the steady increase in the ratio $p$ could eventually trigger a
phase change that would generate particle masses. We are not advocating the
Mannheim model in this paper; it must still meet some serious challenges
before it can be regarded as a serious rival to the conventional model.
We are simply using it as an example of the way the non-thermal component
of the effective mass can play a crucial role.

\begin{figure}
\centering
\includegraphics[scale=0.5]{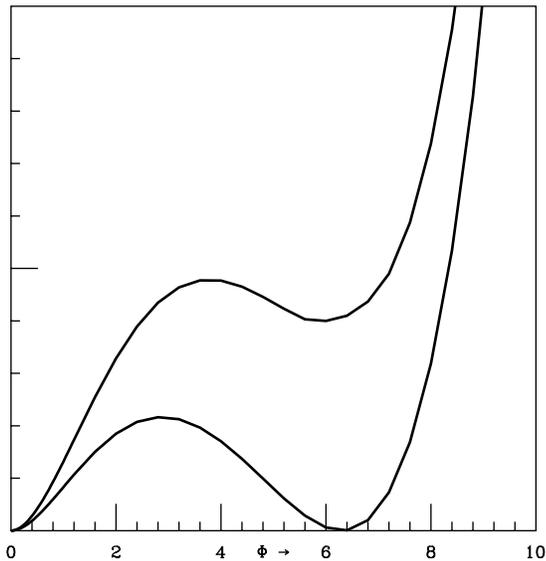}
\caption{Effective potential, $V_{\mathrm{eff}}$, as a function of $\Phi$.
As $m_{\mathrm{nt}}/m_{\mathrm{th}}$ increases, the form of
$V_{\mathrm{eff}}$ changes from the upper curve to the lower, at which
point a Coleman-Weinberg transition can occur. }
\label{fig:Veff}
\end{figure}

The mechanism we have in mind is the Coleman-Weinberg (CW) transition
\citep{cole1}, as discussed in the context of cosmology in \citet{nar2}.
Figure \ref{fig:Veff} shows the effective potential, $V_{\mathrm{eff}}$,
for a typical field theory at finite temperature. (The formulae used are
taken from \citet{nar2}). Initially, $V_{\mathrm{eff}}$ is described by
the upper curve, but as the effective mass of the field quanta increases,
the second minimum will move lower, and eventually, when $p \approx 1$, it
will cross the horizontal axis. At this point a CW transition
can take place and normal masses will appear. These will have
magnitude $m^2 \approx \langle {\mathbf A}^2 \rangle_{\mathrm{nt}}$.

By adjusting the parameters of the model we can arrange for sufficient
conformal time to elapse from the formation of the bubble to the CW
transition, assumed to occur at the weak interaction scale. The initial
temperature of the bubble, $T_{\mathrm{max}}$, is then found to be very
high, and we have to look for an explanation of why particle masses are so
very small compared to this temperature. (This problem is analogous to
the hierarchy problem in ordinary cosmology, where the Planck mass is so
much greater than particle masses.)

In the Mannheim model, the explanation is straightforward. The
temperature of the CW transition is found to depend exponentially on the
conformal time, so a quite modest value of $\eta$ will produce a very small
value of $T_{\mathrm{CW}}/T_{\mathrm{max}}$.

\section{CONCLUSION}
\label{sec:conclude}

We have studied the propagation of the vector potential in the plasma of the
early Universe, assuming a FRW space of negative curvature, and conclude
that $\langle {\mathbf A}^2 \rangle$ will slowly acquire a non-thermal
part, $\langle {\mathbf A}_{\mathrm{nt}}^2 \rangle$, due to distant
matter. Even though this is a direct effect of the vector potential,
and cannot be attributed to fields, the theory is nonetheless gauge
invariant. This effect is therefore in the same category as the
Aharonov-Bohm effect \citep{ahar,tono}.

Further research will be needed to determine whether this increase of
$\langle {\mathbf A}_{\mathrm{nt}}^2 \rangle$ is sufficient to cause a CW
transition and generate a mass scale.

Several additional questions are raised by the present paper, among them the
following:
\begin{enumerate}
	\item Will a CW transition necessarily take place as 
	$\langle {\mathbf A}^2 \rangle_{\mathrm{nt}}$ approaches
	$\langle {\mathbf A}^2 \rangle_{\mathrm{th}}$~?
	\item Is the tail of ${\mathbf A}$ important only in the early Universe,
	or are there other places, regions of high gravitational fields, where
	its effects can be observed even now?
	\item How does the real complicated early plasma determine such things
	as the collision time?
	\item Can we expect ${\mathbf A}$ to propagate over cosmological
	distances so that the CW transition takes place within the available
	conformal time? We have taken findings from ordinary plasma theory and
	used them in the very different circumstances of the early Universe.
\end{enumerate}

These considerations are beyond the scope of this paper, which is solely
concerned with the interplay, in a simple model, of field theory
(density matrix, the Lagrangian for a scalar field, the Coleman-Weinberg
transition) and the classical equations of propagation of the ordinary
vector potential in a FRW space of negative curvature.

A well-known text \citep{pesk1} suggests a connection between large and
small scales similar to the one explored here. After surveying the
difficulties faced by current theories of the mass scale, the authors
write: ``\ldots it may be that the overall scale of energy-momentum is
genuinely ambiguous and is set by a cosmological boundary condition.''
We have presented a mechanism for such a connection. It is based on the
familiar electromagnetic interaction, and nothing radically new seems to be
required.

\section*{Acknowledgments}

We acknowledge helpful correspondence with Bryce DeWitt, Leonard Parker,
Stephen Fulling, Don Melrose, David Montgomery and Philip Mannheim. We
also wish to thank the chairman and faculty of the department of physics
at Washington University for providing an office and computer support for
a retired colleague. Cosmic space may be infinite, but office space is at
a premium.


\bibliography{msep}

\appendix

\section{Normalization with conductivity}
\label{app:normalize}

In this appendix we will work in the usual spherical polar coordinates, and
make connection with Riemannian coordinates when necessary.

Suppose we have a dipole at the origin, oscillating with time dependence
$\exp (-{\mathrm i} \omega t)$ in the $z$ direction. The surrounding medium
is of uniform conductivity, $\overline{\sigma}$, so that the current density,
${\mathbf j}$, is given by ${\mathbf j} = \overline{\sigma} {\mathbf E}$. We
will analyze this system by imagining a small sphere of radius $r_1$ cut out
of the medium surrounding the dipole. Induced currents flowing in the
medium will cause surface charges to appear on the sphere, and the total
dipole moment will be the sum of the original dipole moment and that
due to the induced charges. We assume the permittivity and magnetic
susceptibility are essentially unity, so ${\mathbf D} = {\mathbf E}$ and
${\mathbf B} = {\mathbf H}$. In such a system
$\nabla \cdot {\mathbf j} = 0$ follows from Maxwell's equations, so there
are no volume charges in the medium. 

Denote by $D_{\mathrm{true}} = D (\omega)\exp(-{\mathrm i} \omega t)$
the dipole moment at the center of the sphere, where $D (\omega)$ is
the true dipole strength at angular frequency $\omega$. The induced
dipole moment due to the surface charges is $D_{\mathrm{ind}}$, so the total
dipole moment is $D_{\mathrm{tot}} = D_{\mathrm{true}} + D_{\mathrm{ind}}$.

Just outside the sphere the electrostatic potential is given by
\begin{equation}
V = D_{\mathrm{tot}}P_1 (\cos \theta )/r_1^2 ,
\label{eq:Phi_B}
\end{equation}
where $P_1 (\cos \theta ) = \cos \theta$.
The radial component of ${\mathbf E}$ is given by 
\begin{equation}
E_r = 2D_{\mathrm{tot}} P_1 (\cos \theta )/r_1^3 .
\label{eq:E_r_B}
\end{equation}

The surface charge density, $s$, obeys the relation
\begin{equation}
\frac{{\mathrm d} s}{{\mathrm d} t} = -j_r \;,
\end{equation}
where $j_r$ is evaluated just outside the sphere. This gives
\begin{equation}
s = \frac{-{\mathrm i} \overline{\sigma} E_r}{\omega}
= s_0 P_1 (\cos \theta) ,
\end{equation}
where
\begin{equation}
s_0 = \frac{-2{\mathrm i} \overline{\sigma} D_{\mathrm{tot}}}{r_1^3 \omega} .
\label{eq:s_0}
\end{equation}

The induced dipole moment, $D_{\mathrm{ind}}$, is then given by an integral
over the surface of the sphere:
\begin{eqnarray}
D_{\mathrm{ind}} & = & 2 \pi \int_{0}^{\pi} {\mathrm d}  \theta
\sin \theta r_1^2 [s_0 P_1 (\cos \theta)][r_1^2 P_1 (\cos \theta)]
\nonumber \\
  & = & \left(\frac{-4\pi {\mathrm i}  \overline{\sigma}}{\omega}\right)
\left( \frac{2}{3} \right) D_{\mathrm{tot}}  ,
\end{eqnarray}
giving
\begin{eqnarray}
D_{\mathrm{tot}} & = & D_{\mathrm{true}} + D_{\mathrm{ind}} \nonumber \\
  & = & \frac{3 \omega D_{\mathrm{true}}}
{3 \omega + 8\pi {\mathrm i}  \overline{\sigma}}  .
\end{eqnarray}
 
We set $H_\phi = C_5 (\omega,\overline{\sigma}) N_1 (\theta) h(r)
\exp (-{\mathrm i}  \omega t)$, where $C_5 (\omega,\overline{\sigma})$ is a
normalizing function,
$N_1 = {\mathrm d} P_1 /{\mathrm d} \theta$, and $h(r)$ satisfies
\begin{equation}
\frac{{\mathrm d} ^2 h}{{\mathrm d}  r^2} + \frac{2}{r}
\frac{{\mathrm d} h}{{\mathrm d} r} - \frac{2h}{r^2} + \kappa^2h = 0 \;,
\end{equation}
with $\kappa^2 = \omega (\omega + 4\pi {\mathrm i}  \overline{\sigma})$.
We choose the solution that represents outgoing waves, so
\begin{eqnarray}
h(r) & = & h_1^{(1)} (\kappa r) \nonumber \\
  & = & \left( - \frac{{\mathrm i} }{(\kappa r)^2}
- \frac{1}{\kappa r} \right) \exp ({{\mathrm i} }\kappa r) .
\end{eqnarray}
Here $h_1^{(1)} (\kappa r)$ is the spherical Bessel function defined in
\citet{absteg}.

The Maxwell equation 
$\nabla\times {\mathbf H} =
-{\mathrm i} (\omega + 4\pi {\mathrm i} \overline{\sigma}){\mathbf E}$
then gives
\begin{equation}
E_r = \frac{-2{\mathrm i} C_5 (\omega,\overline{\sigma})
P_1(\cos \theta) h_1^{(1)} (\kappa r) \exp (-{\mathrm i}  \omega t)}
{r (\omega + 4\pi {\mathrm i} \overline{\sigma})} .
\end{equation}
For small $r$ this becomes:
\begin{equation}
E_r = \frac{-2 C_5 (\omega,\overline{\sigma})
P_1(\cos \theta) \exp (-{\mathrm i}  \omega t)}
{\kappa^2 r^3 (\omega + 4\pi {\mathrm i}  \overline{\sigma})} .
\end{equation}

But also, for small $r$, we have (\ref{eq:E_r_B}), so
\begin{eqnarray}
C_5 (\omega,\overline{\sigma})  & = & \frac
{-3 \omega \kappa^2 (\omega + 4\pi {\mathrm i}  \overline{\sigma})
D (\omega) } {(3\omega + 8\pi {\mathrm i}  \overline{\sigma})} ,
\label{eq:C3norm} \\
H_{\phi} & = & \nonumber \\
 & & \hspace{-5em} \frac{-3\omega \kappa^2
(\omega + 4\pi {\mathrm i}  \overline{\sigma}) D (\omega)
N_1 (\theta) h_1^{(1)} (\kappa r) \exp (-{\mathrm i}  \omega t)}
{(3\omega + 8\pi {\mathrm i}  \overline{\sigma})} , \label{eq:H_B} \\
E_r  & = & \frac{6 {\mathrm i}  \omega \kappa^2 D (\omega)
P_1 (\theta) h_1^{(1)} (\kappa r) \exp (-{\mathrm i}  \omega t)}
{ r (3\omega + 8 \pi {\mathrm i}  \overline{\sigma})} .
\label{eq:E_B}
\end{eqnarray}

In this appendix we have used ordinary polar coordinates in flat space.
We need now to transform to coordinates of a FRW flat space, with metric
\begin{equation}
{\mathrm d} s^2 = R^2 (\eta)\left(-{\mathrm d}  \eta^2 +
{\mathrm d} \chi^2 + \chi^2 \, {\mathrm d} \theta^2
+ \chi^2 \sin^2 \theta \, {\mathrm d} \phi^2 \right) .
\label{eq:ds4}
\end{equation}

To get from $H_{\phi}$ of (\ref{eq:H_B}) to $F_{12}$ of (\ref{eq:F12def}),
in the flat FRW metric (\ref{eq:ds4}), we first refer to
\citet{wein2}, section 4.8, and multiply by $\chi$. We then
convert from $t$ to $\eta$ and $r$ to $\chi$ by multiplying by
$R^2 (\eta)$. We also convert $\omega$ to $n$, $\kappa$ to $k$ and
$\overline{\sigma}$ to $\sigma = R(\eta) \overline{\sigma}$:
\begin{eqnarray}
F_{12} & = & \nonumber \\
 & & \hspace{-4em}  \frac
{-3 n k^2 (n + 4\pi {\mathrm i}  \sigma) D (n) N_1 (\theta)
\chi h_1^{(1)} (k \chi) \exp (-{\mathrm i}  n \eta)}
{R(\eta) (3n + 8 \pi {\mathrm i}  \sigma)} .  \label{eq:F12flat}
\end{eqnarray}

We can obtain the corresponding function in curved space by 
multiplying (\ref{eq:f3out2perm}) by
$N_1 (\theta) \exp(-{\mathrm i} n \eta)$:
\begin{eqnarray}
F_{12} & = & \nonumber \\
 & & \hspace{-3em} \frac{C_3 (n,\sigma) N_1 (\theta)
\exp [{\mathrm i} (k \chi - n \eta)]}
{2 u} \left(1 - 2{\mathrm i} ku + u^2 \right) . \label{eq:F12curv}
\end{eqnarray}

We can now find $C_3 (n,\sigma)$ by matching (\ref{eq:F12flat}) with 
(\ref{eq:F12curv}) for small $\chi$:
\begin{equation}
C_3 (n,\sigma) = \frac{3{\mathrm i} n (n + 4\pi {\mathrm i}  \sigma) D (n)}
{R(\eta) (3n + 8 \pi {\mathrm i}  \sigma)} .
\end{equation}

With $D (n)$ given by (\ref{eq:dipstrength}), this reduces to
\begin{equation}
C_3 (n,\sigma) = \frac{-6qV (n + 4\pi {\mathrm i}  \sigma) \sin (n \tau)}
{n (3n + 8 \pi {\mathrm i}  \sigma)} .
\label{eq:normC3perm2}
\end{equation}
In the limit $\sigma \rightarrow 0$, (\ref{eq:normC3perm2}) tends to
(\ref{eq:normC1}), as it should.

\section{The modified Lorenz \\
condition in conformally \\
invariant theories}
\label{app:conform}

It is well known that although Maxwell's equations are conformally invariant
in 4-dimensional space, the Lorenz condition is not
\citep[][chapter 2]{hoyle}. So a theory that aspires to be completely
conformally invariant, such as the theory of Mannheim discussed in section
\ref{sec:mannheim_model}, must include a modified gauge condition if it is
to describe the propagation of potentials. We note that the conformal
weights of $g^{\mu \nu}$, $\sqrt{g}$, $\Phi$ and $A_{\nu}$ are $-2$, $4$,
$-1$ and zero, respectively. The equation
\begin{equation}
\frac{\partial}{\partial x^{\mu}}
\left( \sqrt{g} g^{\mu \nu} \Phi^{*} \Phi A_{\nu} \right) = 0
\end{equation}
is therefore conformally invariant, and is a candidate for a gauge
condition. We can linearize it by setting $\Phi^{*} \Phi$ equal to its
expectation value. In the conditions of the early Universe, this will be
proportional to $T^2$, where $T$ is the temperature. Assuming adiabatic
expansion, $T^2$ will be proportional to $1/R^2$, so our gauge condition
reduces to
\begin{equation}
\frac{\partial}{\partial x^{\mu}}
\left( \sqrt{g} g^{\mu \nu} A_{\nu}/R^2 \right) = 0 ,
\end{equation}
which is the modified Lorenz condition, (\ref{eq:lrnz2}).

\bsp

\end{document}